\documentclass[lettersize,journal]{IEEEtran}
\usepackage[utf8]{inputenc}  
\usepackage[utf8]{inputenc} 
\usepackage[T1]{fontenc}

\usepackage{amsmath,amsfonts}
\usepackage{algorithmic}
\usepackage{algorithm}
\usepackage{array}
\usepackage[caption=false,font=normalsize,labelfont=sf,textfont=sf]{subfig}
\usepackage{textcomp}
\usepackage{stfloats}

\usepackage{setspace} 
\usepackage{url}
\usepackage{verbatim}
\usepackage[none]{hyphenat} 
\usepackage{graphicx}
\usepackage{cite}
\begin{document}

\title{Efficient Near-Field Beam Focusing Merging Orthogonal Matching Pursuit and CVX for Large Intelligent Surface Applications }

\author{Jiawang Li,~\IEEEmembership{Student Member,~IEEE}
\thanks{Manuscript received xxxx.xx.xx. (Corresponding author: \textit{Jiawang Li})}
\thanks{Jiawang Li is with the Department of Electrical and 
Information Technology, Lund University, 22100 Lund, Sweden 
(e-mail: {jiawang.li@eit.lth.se}).}
\thanks{Color versions of one or more of the figures in this letter are available online at http://ieeexplore.ieee.org.}
\thanks{Digital Object Identifier 10.1109/AWPL.2025.xxx.}}

\markboth{Journal of \LaTeX\ Class Files,~Vol.~14, No.~8, February~2025}%
{Shell \MakeLowercase{\textit{et al.}}: A Sample Article Using IEEEtran.cls for IEEE Journals}

\maketitle

\begin{abstract}
In this paper, an efficient near-field beamforming method is proposed to support the large intelligent surfaces (LIS) that are expected to be widely deployed in 6G networks. This approach avoids directly applying convex (CVX) optimization for sparse selection in large-size array matrices, as such methods often lead to excessive computational time due to blind searching to satisfy a series of objective functions. First, based on the objective function, we prioritize a key component and employ the orthogonal matching pursuit (OMP) method to pre-select potential sparse target positions. To ensure focal symmetry, a coordinate mirror symmetry approach is adopted, meaning that selection is performed only in the first quadrant, while the remaining quadrants are determined through mirror symmetry relative to the first quadrant. This significantly reduces computational complexity at an early stage. Next, CVX is applied based on the pre-selected sparse array. Once a predefined threshold is met, a solution is obtained that satisfies the constraints of the beamfocusing. 
The results demonstrate that, compared with conventional methods, this approach improves efficiency by 15.12 times with 121 elements and 96.73 times with 441 elements. The proposed method demonstrates not only satisfactory performance but also considerable potential as a beam focusing technique for large-scale near-field array systems.

\end{abstract}


\section{Introduction}
\IEEEPARstart{I}{n} recent years, near-field focusing (NFF) technology has garnered increasing attention due to its wide range of applications. This technique enables energy from an array to be concentrated within a specific spatial region, making it highly suitable for applications such as wireless communication [1]-[3], power transfer [4]-[6] and microwave imaging [7]-[8]. Unlike far-field beamforming methods [9], near-field techniques predominantly employ phase-conjugation, which offers clearer physical interpretability and greater robustness. However, most existing research has focused primarily on array structures, often resulting in high sidelobe levels in the beam pattern. In practical applications, such as communication, where interference must be minimized, and power transmission, where efficiency needs to be maximized—a low-sidelobe focal point design is essential.

Large intelligent surfaces (LIS), recognized as a promising technology for beyond-5G (B5G) and early 6G mobile communications, are typically composed of hundreds or even thousands of elements [10]. Selecting a subset of these elements to generate a low-sidelobe focal point presents a significant computational challenge [11]-[13]. such as convex (CVX) optimization-based Bayesian compressive sensing [11], group $\ell_{1/2}$ regularization [12] and reweighted $\ell_1$-norm optimization [13] are proposed to synthesize near-field focusing sparse phased arrays (NFF SPAs) with a minimal number of elements. In indoor environments, where users are mobile, beam focusing must dynamically track user trajectories, imposing stringent real-time adaptation requirements on LIS. Additionally, in scenarios where users transition from distant locations to closer proximity, dynamic power adjustment is essential to ensure optimal focusing. However, this does not imply that minimizing the number of activated elements is always beneficial for energy transmission between the LIS and the user. Although utilizing fewer elements can enhance energy efficiency and reduce design complexity, an overly sparse selection may compromise performance in certain applications. While these methods [11]-[13] are practical and ensure convergence, their performance drops significantly.

A high-efficiency near-field beam focusing method is proposed for LIS in 6G networks. This approach avoids direct CVX optimization for sparse selection in large-size arrays, which is computationally expensive due to blind searching.
First, based on the objective function, a key component is prioritized, and the orthogonal matching pursuit (OMP) method is used to pre-select sparse target positions. To ensure focal symmetry, selection is performed only in the first quadrant, with the rest determined by mirror symmetry, significantly reducing complexity.
Next, CVX method is applied to the pre-selected array. Once a predefined threshold is met, a beamforming solution is obtained.
Compared to traditional methods, this approach improves efficiency  by 15.12 times with 121 elements and 96.73 times
with 441 elements.

\section{Algorithm Description}
\subsection{Problem Formulation}
A planar Near-Field Focused (NFF) array consisting of \( N \) arbitrarily positioned elements is considered, where each element is located at \( \vec{r}_n = (x_n, y_n, z_n) \) for \( n = 1, 2, \dots, N \). The total electric field at an arbitrary observation point \( \vec{r} = (x, y, z) \) in the near-field region can be expressed as

\begin{equation}
    E(\vec{r}) = \sum_{n=1}^{N} w_n f_n(\vec{r}) \frac{e^{-j k |\vec{r} - \vec{r}_n|}}{|\vec{r} - \vec{r}_n|}
\end{equation}

where \( w_n \in \mathbb{C} \) denotes the complex excitation coefficient of the \( n \)-th element, \( f_n(\vec{r}) \) represents the copolarization component of the element’s radiation pattern, and \( k = \frac{2\pi}{\lambda} \) is the wave number in free space, with \( \lambda \) denoting the wavelength. The observation point is located within the Fresnel zone, where near-field effects significantly influence beam focusing. To make the elements keep the same polarization, a \( y \)-oriented dipole is adopted as the radiating element. The objective is to determine the sparse set of nonzero excitations that can produce the desired radiation pattern in a very short time. 
\begin{align}
    \text{s.t.} \quad 
    & A(0,0,z_0) \cdot w = E_{max},  \quad \text{}  \nonumber \\
    & \left| A(x_s,y_s,z_0) \cdot w \right| \leq A_{SL}, \quad \text{}  \nonumber \\
    & \left| A(0,0,z) \cdot w \right| \leq E_{max}, \quad \text{}
\end{align}
where $E_{max}$ is the peak value of the observation plane, it can be normalized to 1, which means $\frac{|A(x, y, z_0) \cdot w|}{|A(0, 0, z_0) \cdot w|} \leq 1$. $A_{SL}$ is the preset maximum sidelobe level (SLL) is within the observation area. $x_s$ and $y_s$ is the sampling points apart from the mainlobe at $z$ = $z_0$ plane. $A(\cdot)$ is the $m \times n$ signal propagation matrix, which can be described as: 
\begin{equation}
{A} =
\begin{pmatrix}
e^{j k |r_1 - r_1|} & \cdots & e^{j k |r_1 - r_n|} \\
\vdots & \ddots & \vdots \\
e^{j |r_m - r_1|} & \cdots & e^{j k |r_m - r_n|}
\end{pmatrix}
\end{equation}
$m$ is the number of the sample points and $n$ is the number of the antenna elements.
\subsection{Algorithm Process}
Although conventional $l_1$-norm CVX optimization methods can achieve a sparse antenna array configuration with fewer elements while meeting design requirements, this comes at the cost of limited flexibility in adjusting the focal intensity [13]. Consequently, in many practical scenarios, obtaining the sparsest solution is not the primary objective. Instead, rapid convergence is often more critical, particularly in dynamic environments where users are in motion. In such cases, spending more than 100s to shift the focal point to the next position is impractical. This challenge becomes even more pronounced in large-scale LIS applications, where the increasing matrix dimensions inevitably lead to the computational intractability of conventional search methods.

To address this issue, we propose a joint search method. Firstly, the OMP method is employed to identify the optimal positions and excitations of target antenna elements that maximize the amplitude at the focal point. In the \( k \)-th iteration, we first compute the correlation between all columns of the sensing matrix \( A \) and the current residual \( \mathbf{r}^{(k)} \):
\begin{equation}
c_j = \left| \mathbf{a}_j^T \mathbf{r}^{(k)} \right|
\end{equation}
\begin{algorithm}[H]
\caption{The proposed method}
\label{alg:svd_krosbl}
\begin{algorithmic}[1]
\REQUIRE Field matrix $A$, wavelength $\lambda$, sidelobe constraint $\rho_{SL}$, max sparsity $s_{\max}$, tolerance $\varepsilon$
\STATE \textbf{Initialization:} residual $r \leftarrow$ target field; active set $\mathcal{S} \leftarrow \emptyset$
\FOR{$\text{iter} = 1$ to $s_{\max}$}
    \STATE Compute correlation vector $c \leftarrow |A^H r|$
    \STATE Exclude selected elements: $c(\mathcal{S}) \leftarrow 0$
    \STATE Select index $i^* \leftarrow \arg\max c$
    \STATE Find symmetric counterparts of $i^*$ and update $\mathcal{S}$
    \STATE Construct $A_{\mathcal{S}}$ using selected elements
    \STATE Solve least squares: $w_{\mathcal{S}} \leftarrow A_{\mathcal{S}}^\dagger \cdot$ target
    \STATE Update residual: $r \leftarrow$ target $- A_{\mathcal{S}} w_{\mathcal{S}}$
    \IF{$\|r\|_2 < \varepsilon$}
        \STATE \textbf{break}
    \ENDIF
\ENDFOR
\STATE \textbf{CVX Optimization:}
\STATE Minimize $\|w_{\mathcal{S}}\|_1$ subject to:
\STATE \quad Mainlobe constraint: $A_1 w_{\mathcal{S}} = 1$
\STATE \quad Sidelobe constraint: $|A_{SL} w_{\mathcal{S}}| \leq \rho_{SL}$
\STATE \quad Axial constraint: $|A_Z w_{\mathcal{S}}| \leq 1$
\STATE \textbf{Result:} Output optimized sparse weights $w_{\mathcal{S}}$
\end{algorithmic}
\end{algorithm}
where \( c_j \) represents the correlation of the \( j \)-th atom. After selecting the index \( j^* \) corresponding to the highest correlation, its symmetric indices are further included to ensure the structural symmetry of the beam.
The selected atom set \( \Lambda^{(k)} \) forms the submatrix \( A_{\Lambda^{(k)}} \), and its sparse excitation is obtained by solving the least squares problem:

\begin{equation}
\mathbf{w}^{(k)} = A_{\Lambda^{(k)}}^{\dagger} \mathbf{y}
\end{equation}
where \( \dagger \) denotes the Moore-Penrose pseudoinverse. The residual is then updated:

\begin{equation}
\mathbf{r}^{(k+1)} = \mathbf{y} - A_{\Lambda^{(k)}} \mathbf{w}^{(k)}
\end{equation}
The algorithm terminates when the stopping criterion is met:

\begin{equation}
\|\mathbf{r}^{(k+1)}\| < \epsilon.
\end{equation}
 The efficiency of the OMP method provides a solid foundation for subsequent convex optimization, enabling the suppression of sidelobes to the required level. To further enhance computational efficiency while ensuring beam symmetry, only one quadrant of the data is processed, and the remaining three quadrants are generated using a mirror symmetry approach. The algorithm flowchart is illustrated in flowchart Algorithm 1.
\subsection{Low SLL design for Small-size Array}
In the initial example, a near-field single focal point pattern with a low SLL, aligned with the approach in [13], is examined. The array's aperture is limited to \(5\lambda \times 5\lambda\), with the focal point positioned at \((0, 0, 5\lambda)\). The initial configuration consists of \(11 \times 11\) dipole elements arranged on the \(xoy\)-plane with an element spacing of \(d_c = 0.5\lambda\). The sampling area on the focal-plane \( xoy\) is set to match the physical aperture of the array, with 50 sampling points uniformly distributed along each side. 
\begin{figure}[H]
    \centering
\includegraphics[width=1\linewidth]{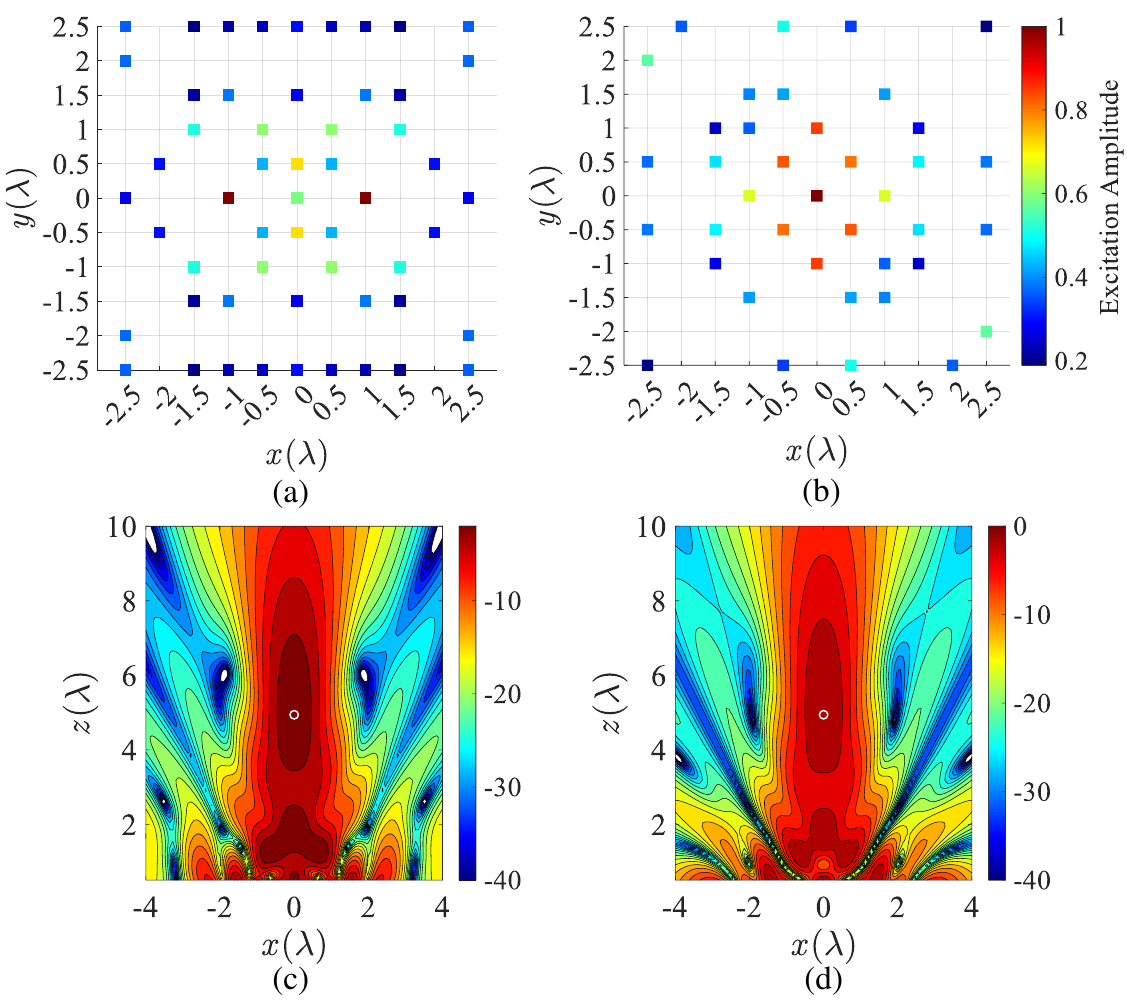} 
    \caption{Sparse array layout and E-field distribution in $\it{xoz}$ plane. (a) Proposed sparse array layout. (b) Conventional sparse array. (c) Proposed E-field distribution in $\it{xoz}$ plane. (d) Conventional E-field distribution in $\it{xoz}$ plane.}
    \label{fig:example}
    \centering
\includegraphics[width=0.95\linewidth]{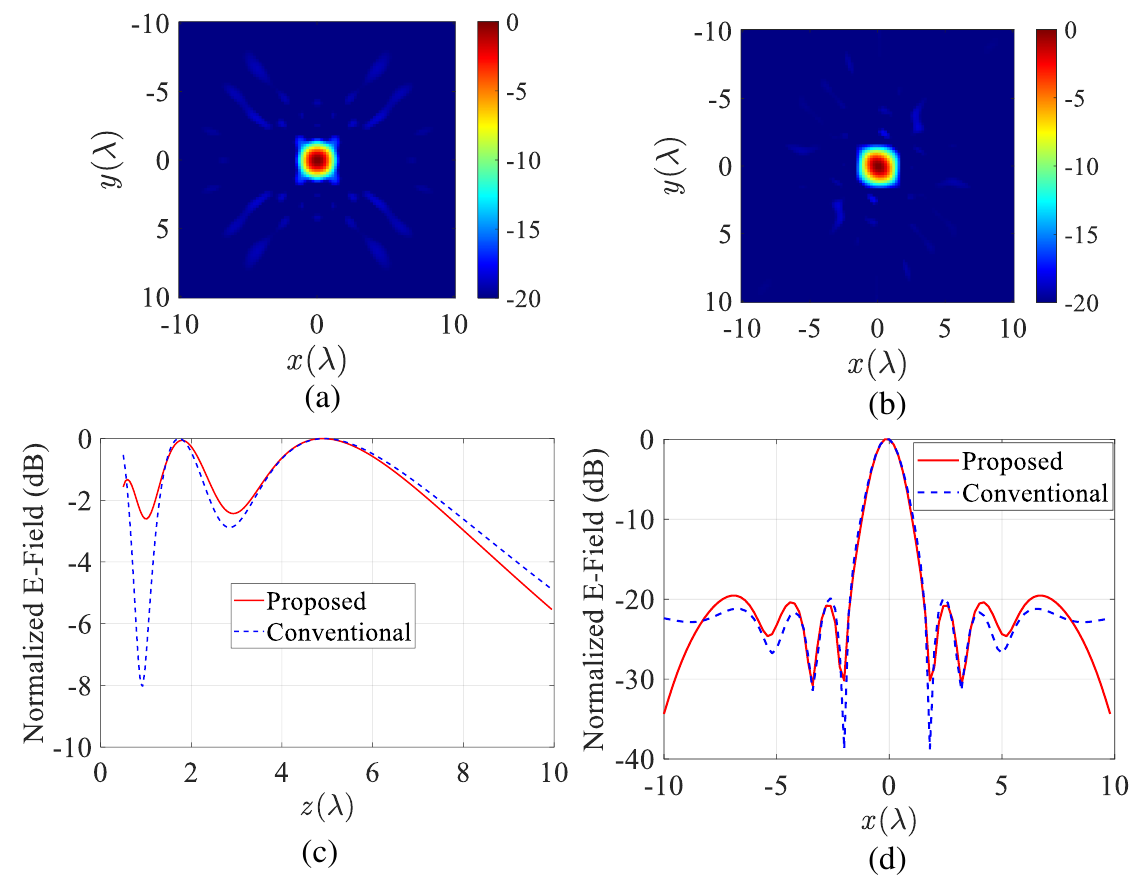} 
    \caption{E-field distribution. (a) Proposed E-field distribution in $\it{xoy}$ plane. (b) Conventional E-field distribution in $\it{xoy}$ plane. (c) 2D-cut E-field along the $\it{z}$-axis. (d) 2D-cut E-field along the $\it{x}$-axis.}
    \label{fig:example}
\end{figure}
Additionally, 100 sampling points are placed along the central line of the \( xoz \) plane to evaluate the axial response. The desired 3 dB beamwidth (BW) and SLL are constrained to \(1.08\lambda\) and \(-20\) dB, respectively.
The threshold \(\varepsilon\) is set to 0.001. The near field region surrounding the array is confined to 
\( x \in [-10\lambda, 10\lambda] \), 
\( y \in [-10\lambda, 10\lambda] \), and 
\( z \in [0.1\lambda, 10\lambda] \).
The synthesized sparse array layout and corresponding E-filed distribution are illustrated in Fig. 1 and Fig. 2. From Fig.~1(a) and (b), it can be observed that the proposed design yields a total of 55 elements with a fully symmetric array structure, whereas the conventional method produces 39 elements with a more irregular spatial distribution. Both configurations exhibit a concentration of elements with higher excitation amplitudes near the central region.
From Fig.~1(c) and (d), where the white circles indicate the beam peaks, it is evident that the proposed method produces a narrower main lobe in both beam width and beam depth, indicating superior spatial resolution capability.
In Fig.~2(a) and (b), the focal fields generated by the proposed method appear symmetrically distributed due to the array symmetry, while the conventional method exhibits a beam stretching effect.
Additionally, the 2D cross-sectional views in Fig.~2(c) and (d) show that the conventional design results in slightly elevated sidelobes near the main lobe (just above $-20$~dB), whereas in the proposed design, SLLs exceeding $-20$~dB occur farther from the main beam—beneficial for mitigating interference signals.
To more clearly demonstrate the advantages of the proposed method in near-field beam focusing, 
key performance metrics—including sparsity, beamwidth, SLL, average peak field strength (Normalized to each antenna element), 
focal shift, and runtime—are summarized in Table \uppercase\expandafter{\romannumeral 1}.
\begin{table}[h]
    \centering
    \scriptsize 
    \caption{Performance Comparison for the NFF Pattern}
    \label{tab:performance_comparison}
    \setlength{\tabcolsep}{2pt} 
    \begin{tabular}{|c|c|c|c|c|c|c|}
        \hline
        \hline
        & \textbf{\boldmath$S(\%)$} & \textbf{BW/\boldmath$\lambda$} & \textbf{SLL (dB)} & \textbf{\boldmath$|E_p|$ (V/m)} & \textbf{\boldmath$\Delta z/\lambda$} & \textbf{Time (s)} \\
        \hline
        CVX-based [13] &  32.2 & 1.236 & -19.93 & 59.72 & 0.069 & 27.4085 \\
        \hline
        Proposed & 45.5 & 1.212 & -19.62 & 55.29 & 0.060 & 1.8123 \\
        \hline
    \end{tabular}
\end{table}
As shown in Table \uppercase\expandafter{\romannumeral 1}, while maintaining comparable performance in terms of sparsity, main beamwidth, SLL, electric field intensity, and focal shift, the proposed design significantly reduces the runtime from 27.4085 s to just 1.8123 s.
\subsection{Low SLL design for Large-size Array}
In the next step, we conduct validations on ultra-large-scale arrays, which hold great potential for applications in the millimeter-wave and even THz bands. In this simulation, the array size is expanded to $21 \times 21$, and the simulation results are statistically analyzed. The number of sampling points in the $xoy$ plane is set to 100 per side, while 200 sampling points are configured along the $z$-axis. The focal point remains at $(0, 0, 5\lambda)$. The corresponding simulation results are shown in Fig. 3 and Fig. 4. As illustrated in Fig. 3, in addition to the conventional concentration of active elements near the array center, this design also features distributed and weakly excited elements near the array edges. From Fig. 4(c) and (d), it is evident that this edge-distributed configuration improves the focusing resolution, particularly enhancing the focal depth. In contrast, traditional methods exhibit a shifted peak electric field location, approximately at $3\lambda$, along with an elongated focal depth region, as also observed in Fig. 4(c). Furthermore, the asymmetry in the amplitude distribution of the traditional approach leads to asymmetric sidelobes in the focal region, which may result in locally elevated sidelobes. A comparison between Fig. 4(a) and (b) further reveals that the proposed design achieves a more concentrated focal spot and significantly reduced sidelobe levels around it, thereby greatly mitigating interference with surrounding targets in practical applications.
Furthermore, in this step, the obtained amplitude and phase distributions are imported into the full-wave electromagnetic simulation software \textsc{FEKO}~2023 to validate the simulation results shown in Fig. 4(c) and Fig. 4(d).
\begin{figure}[H]
    \centering
\includegraphics[width=1.05\linewidth]{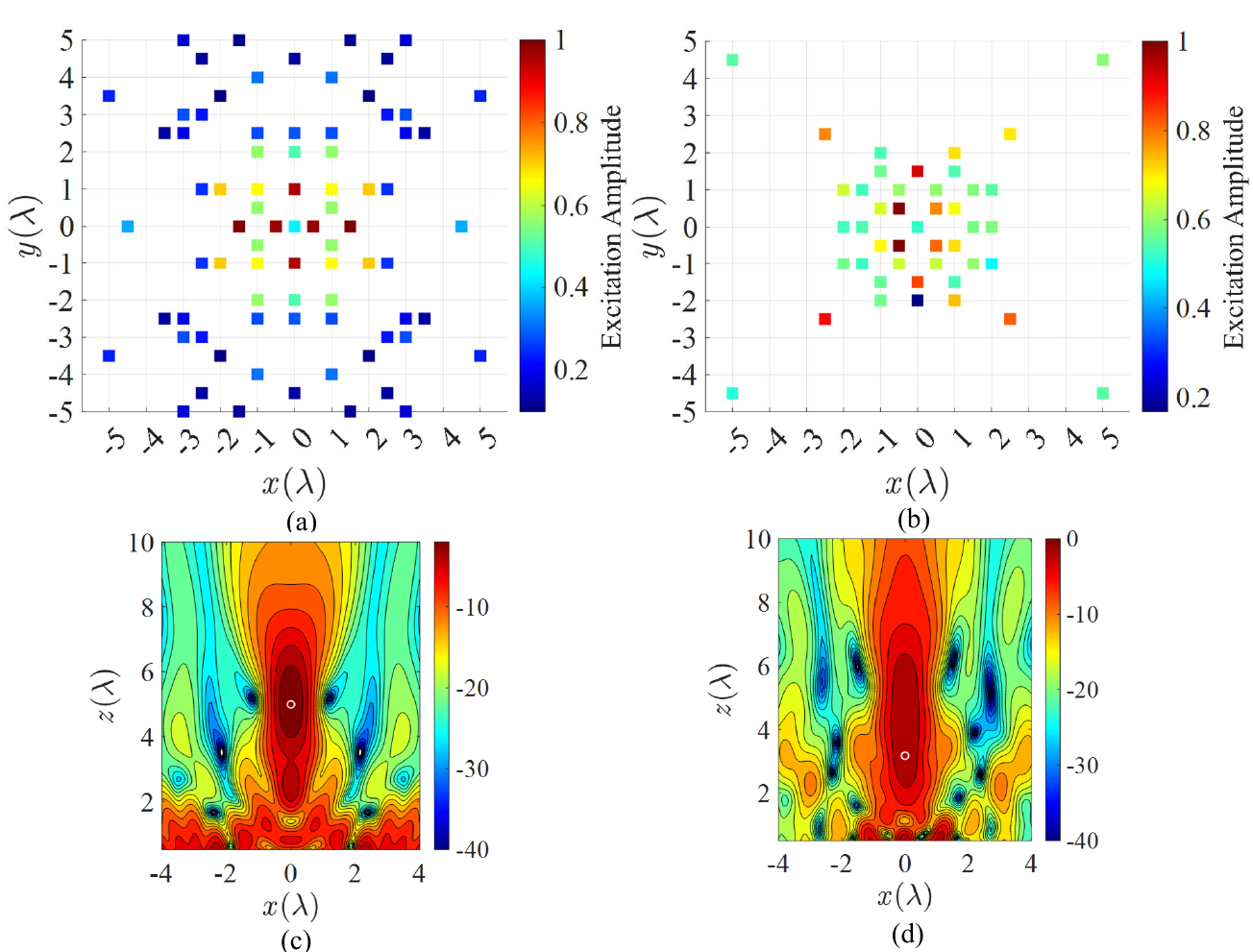} 
    \caption{Sparse array layout and E-field distribution in $\it{xoz}$ plane. (a) Proposed sparse array layout. (b) Conventional sparse array. (c) Proposed E-field distribution in $\it{xoz}$ plane. (d) Conventional E-field distribution in $\it{xoz}$ plane.}
    \label{fig:example}
\end{figure}
\begin{figure}[H]
    \centering
\includegraphics[width=1\linewidth]{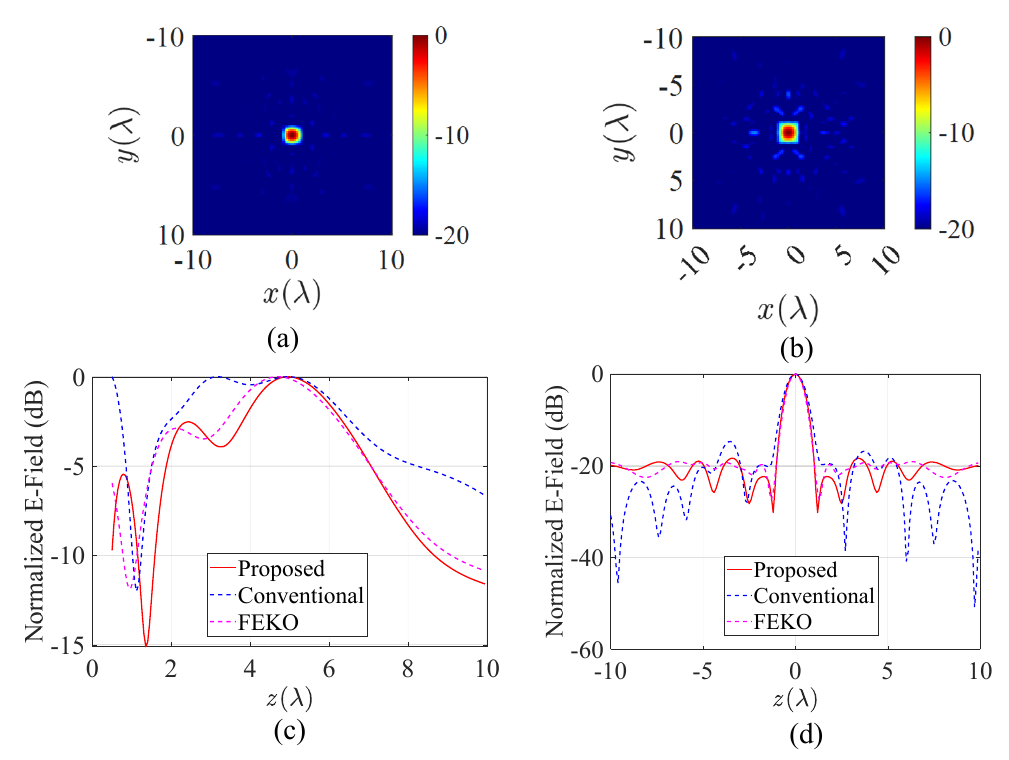} 
    \caption{E-field distribution. (a) Proposed E-field distribution in $\it{xoy}$ plane. (b) Conventional E-field distribution in $\it{xoy}$ plane. (c) 2D-cut E-field along the $\it{z}$-axis. (d) 2D-cut E-field along the $\it{x}$-axis.}
    \label{fig:example}
\end{figure}
Table \uppercase\expandafter{\romannumeral 2} summarizes the near-field beam synthesis metrics for the large-scale array. Compared with the traditional method, the 3\,dB beamwidth is reduced from $1.02\lambda$ to $0.90\lambda$, indicating an improvement in lateral resolution. Meanwhile, the sidelobe level improves from a suboptimal --14.70\,dB to --18.35\,dB. However, the most significant advantage of the proposed design lies in computational efficiency: the execution time is drastically reduced from 715.39 s to only 7.40 s, an improvement of approximately 96.7 times. This substantial acceleration in convergence highlights the potential of the method for real-time near-field beam tracking applications.

\begin{figure}[H]
    \centering
\includegraphics[width=0.75\linewidth]{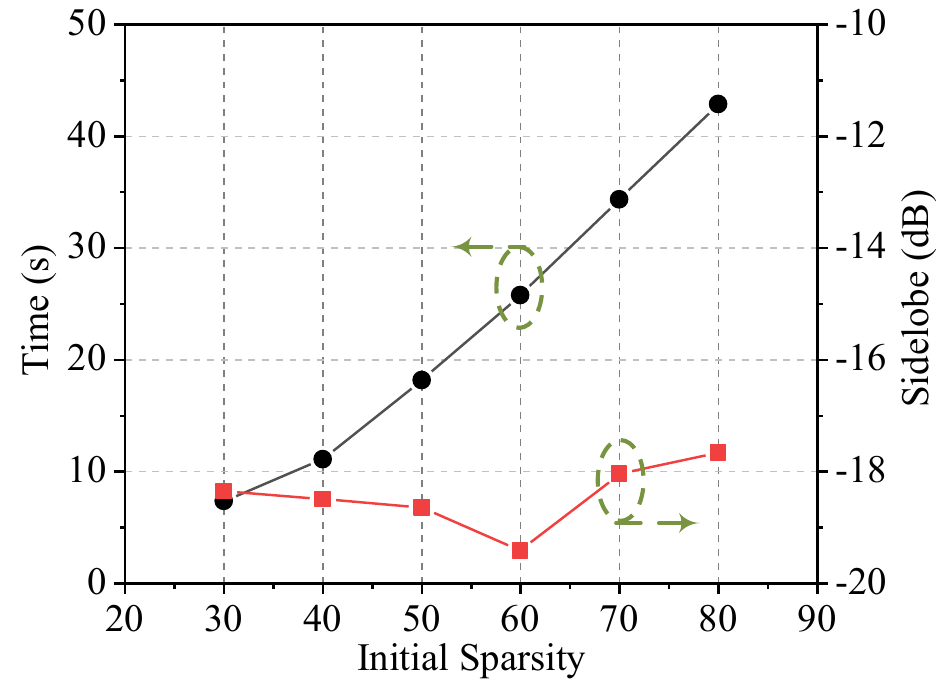} 
    \caption{Relationship between the initial sparsity setting, synthesized radiation pattern sidelobe level, and computational time.}
    \label{fig:example}
\end{figure}
\begin{table}[h]
    \centering
    \scriptsize 
    \caption{Performance Comparison for the NFF Pattern}
    \label{tab:performance_comparison}
    \setlength{\tabcolsep}{2pt} 
    \begin{tabular}{|c|c|c|c|c|c|c|}
        \hline
        \hline
        & \textbf{\boldmath$S(\%)$} & \textbf{BW/\boldmath$\lambda$} & \textbf{SLL (dB)} & \textbf{\boldmath$|E_p|$ (V/m)} & \textbf{\boldmath$\Delta z/\lambda$} & \textbf{Time (s)} \\
        \hline
        CVX-based [13] &  9.98 & 1.020 & -14.6967 & 95.83 & 0.013 & 715.3932 \\
        \hline
        Proposed & 17.9 & 0.900 & -18.35 & 97.13 & 0.013 & 7.3954 \\
        \hline
    \end{tabular}
\end{table}
In addition, simulations are conducted to evaluate the effect of the initial sparsity setting in the OMP algorithm. The simulation also records the time required to complete each iteration and visualizes the corresponding sidelobe levels. The results are illustrated in Fig. 5. Here, the initial sparsity refers to a single quadrant in the Cartesian coordinate system; therefore, the actual specified sparsity should be multiplied by a factor of four. As shown in the results, the execution time is strongly correlated with the initial sparsity setting. For typical indoor LIS applications, choosing a smaller initial sparsity facilitates faster convergence of beam focusing. Regarding the synthesized SLLs, the results remain relatively stable at approximately 18.5\,dB. Although this is slightly below the expected level of -20\,dB, further SLL suppression can be achieved through additional windowing techniques.
\section{Conclusion}
In this paper, an efficient near-field beamforming method is proposed to support the LIS that are expected to be widely deployed in 6G networks. This approach avoids directly applying convex optimization for sparse selection in high-dimensional array matrices, as such methods often lead to excessive computational time due to blind searching to satisfy a series of objective functions. First, based on the objective function, we prioritize a important component and employ the OMP method to pre-select potential sparse target positions. To ensure focal symmetry, a coordinate mirror symmetry approach is adopted, meaning that selection is performed only in the first quadrant, while the remaining quadrants are determined through mirror symmetry relative to the first quadrant. This significantly reduces computational complexity at an early stage. Next, CVX is applied based on the pre-selected sparse array. Once a predefined threshold is met, a solution satisfying the beam focusing constraints is obtained.

\newpage

\vfill

\end{document}